# Graphene Composites with Dual Functionality: Electromagnetic Shielding and Thermal Management

Fariborz Kargar[§,×], Zahra Barani[§], Michael G. Balinskiy, Andres Sanchez Magana, Jacob S. Lewis, and Alexander A. Balandin[×]

Phonon Optimized Engineered Materials (POEM) Center, Department of Electrical and Computer Engineering, Materials Science and Engineering Program, Bourns College of Engineering, University of California, Riverside, California 92521 USA

[§] Contributed equally to the work.

[×] Corresponding authors (F.K.): fkargar@engr.ucr.edu and (A.A.B.): balandin@ece.ucr.edu ; web-site: http://balandingroup.ucr.edu/

## Abstract

We report on the synthesis and characterization of the epoxy-based composites with the few-layer graphene fillers, which are capable of the duel functional applications. It was found that composites with the certain types of few-layer graphene fillers reveal an efficient total electromagnetic interference shielding, $SE_{tot} \sim 45$ dB, in the important X-band frequency range, $f = 8.2$ GHz $- 12.4$ GHz, while simultaneously providing the high thermal conductivity, $K \approx 8$ Wm$^{-1}$K$^{-1}$, which is a factor of $\times 35$ larger than that of the base matrix material. The efficiency of the dual functional application depends on the filler characteristics: thickness, lateral dimensions, aspect ratio and concentration. Graphene loading fractions above the *percolation threshold* allow for strong enhancement of both the electromagnetic interference shielding and heat conduction properties. Interestingly, graphene composites can block the electromagnetic energy even below the percolation threshold, remaining electrically insulating, which is an important feature for some types of thermal interface materials. The dual functionality of the graphene composites can substantially improve the electromagnetic shielding and thermal management of the airborne systems while simultaneously reducing their weight and cost.

## I. Introduction

Heat and electromagnetic (EM) waves are inevitable byproducts of all electronic devices, particularly those operating at high frequencies. As the electronic devices decrease in size, and operate at ever increasing frequencies, they produce more heat and EM waves, which result in faster degradation of such devices and negative effects on adjacent electronic systems[1–11]. In addition, EM radiation is a major concern for human health and environment[12–14]. The current industrial and safety standards require blocking of more than 99% of the EM radiation from any electronic devices[1,15–17]. From the other side, the operation of the electronic devices can be disrupted by the outside EM waves. The heat and EM radiation have an inherent connection – absorption of EM waves by any material results in its heating. The energy from EM wave transfers to electrons and then to phonons – quanta of crystal lattice vibrations. The conventional approach for handling the heat and EM radiation problems is based on utilization of the thermal interface materials (TIM), which can spread the heat, and electromagnetic interference (EMI) shielding materials, which can protect from EM waves. These two types of materials have different, and, often, opposite characteristics, *e.g.* excellent EMI material can be a poor heat conductor, while efficient TIM can utilize electrically non-conductive fillers, resulting in its transparency for EM waves. Here, we propose a concept of the *dual-functional* EMI shielding – TIM materials, and demonstrate it on the example of graphene composites.

It is well-known that EMI shielding requires interaction of the EM waves with the charge carriers inside the material so that EM radiation is reflected or absorbed. For this reason, the EMI shielding material must be electrically conductive or contain electrically conductive fillers, although a high electrical conductivity is not required. The bulk electrical resistivity on the order of 1 Ωcm is sufficient for most of EMI shielding applications[1,3,15]. Most of the polymer-based materials widely used as TIMs in electronic packaging are electrically insulating and, therefore, transmit EM waves. Conventionally, metal particles are added as fillers in high volume fractions to the base polymer matrix in order to increase the electrical conductivity, and prevent EM wave propagation from the device to the environment and vice versa [1,18–21]. However, the polymer-metal composites suffer from high weight, cost and corrosion, which make them an undesirable choice for the state-of-the-art downscaled electronics. Several studies reported the use of carbon fibers[22–29], carbon black [30,31], bulk graphite[32–34], carbon nanotubes (CNT)[16,17,35–39], reduced graphene oxide (rGO) [2,6,40–50], graphene[51–54] and, combination of carbon allotropes with or without other metallic –or non-metallic particles [40,42,51,53,55–60] as fillers in various composites for EMI shielding purposes. Other types of advanced fillers, which allow for synthesis of composites with the high EMI shielding efficiency include the sodium alginate with two-dimensional transition metal carbide (MXenes)[1]. At the same time, the thermal properties of EMI shielding materials remain rather poor or not explored.

The proposed dual functionality of a material, which can simultaneously spread the heat, *i.e.* serve as TIM, and shield from EM waves may present enormous technological and cost benefits. They become even greater for applications involving the high-power EM waves. Part of the incident EM wave that propagates inside the EMI shielding material turns into heat, as it is absorbed or reflected internally and, thus, increases the temperature of the EMI shielding material. The temperature rise reduces the electrical conductivity and, as a result, decreases the shielding efficiency of the material. The temperature rise is of major concern for electronic devices in high-tech and medical applications. All these factors create string motivations for the development of such dual functional materials. In this paper, we show that properly optimized composites with few-layer graphene (FLG) fillers can efficiently perform two functions – EMI shielding and thermal management – owing to their excellent electrical and thermal properties, as well as excellent dispersion in and coupling to the matrix materials.

Graphene and FLG are good conductors of electricity. The typically reported values for the sheet resistance of SLG and FLG vary from ~100 Ω up to 30 kΩ depending on the number of layers and quality [61–63]. This is required for the fillers used in EMI shielding materials. Graphene also has extremely high thermal conductivity. The reported values of the *intrinsic* thermal conductivity of high-quality large graphene layers is in the range from 2000 $Wm^{-1}K^{-1}$ to 5000 $Wm^{-1}K^{-1}$ near room temperature (RT)[64–66]. The intrinsic thermal conductivity of graphene can exceed that of the high-quality bulk graphite, which by itself is high − 2000 $Wm^{-1}K^{-1}$ at RT[64,67–69]. Numerous studies reported enhancement of the thermal properties of (TIM) and various other composites as a result of incorporation of single-layer graphene (SLG) and FLG[68,70–76]. The first studies showed that adding even a small loading fraction of the optimized mixture of graphene and FLG (up to $f = 10$ vol. %) to the pristine epoxy increases its thermal conductivity by a factor of ×25[68]. Independent, follow up studies, demonstrated even larger enhancement in the thermal transport properties of composites at lower loading fractions (∼5 vol. %) [77,78]. One of the conclusions from the reports of the thermal properties of composites with graphene and FLG is that there exists an optimum range of the filler lateral dimensions, thicknesses and aspect ratios for heat conduction. The FLG filler can perform better than SLG filler even though it has lower intrinsic thermal conductivity[68,72,79]. The latter is due to the fact that the heat conduction properties of FLG experience less degradation upon exposure to the matrix material. From the other side, if the thickness of FLG becomes too large, the mechanical flexibility of the fillers degrades, resulting in weaker coupling to the matrix material. It should be noted that the mechanical flexibility and excellent coupling of graphene to polymer matrix make it more favorable filler material than other carbon allotropes. The scalable and cost-effective production methods of graphene and FLG via liquid phase exfoliation (LPE)[80,81] or reduction of graphene oxide (GO)[82–85] allow for industrial applications of graphene as fillers in composites. For simplicity, below we use the term "graphene" fillers for a mixture of graphene and FLG with the thicknesses from a single atomic plane, *i.e.* 0.35 nm, to tens of nanometers, and lateral dimensions in a few µm range. This allows us to distinguish graphene fillers from carbon black, nm-scale graphite nano-platelets, or milled µm- and mm-scaled graphite particles[86].

Prior studies suggested that the graphene-based composite exhibit the *electrical percolation threshold* at rather low loading fractions of graphene or FLG fillers[87,88]. A possibility of preparing composites with graphene loading exceeding the electrical percolation threshold is important for designing the EMI shielding materials. The two main mechanisms for blocking EM waves involve the reflection of EM waves via interaction of EM field with the charge carriers, and absorption of EM waves via interaction of EM waves with the electric or magnetic dipoles in the material[1,3,15,18]. The third mechanism, which involves multiple internal reflections of EM waves from the surfaces, scattering centers and defects inside the composite is negligible if the absorption contribution, $SE_A$, to the total EMI shielding, $SE_{tot}$, is less than $\sim 10 - 15$ dB[2,37,89] We have recently found that the thermal percolation follows soon after the electrical percolation in graphene composites[79]. Given the importance of electrical percolation for EMI shielding and thermal percolation for TIM applications, we examined the properties of the designer graphene composites over a wide range of the graphene and FLG loading fractions.

## II. Material Synthesis

We utilized commercially available FLG (Graphene Supermarket) to prepare composites with the high loading fraction of fillers. The material was processed in-house to find the optimum aspect ratio, lateral dimensions and thickness of FLG fillers. For EMI shielding applications, it is desirable to have fillers with the high aspect ratios in order to achieve the electrical percolation at lower filler contents. The theory and experimental studies[90–94] suggest that the higher the aspect ratio of the conductive fillers, the lower is the filler concentration required for the electrical percolation. The electrical percolation and resulting electrical conduction via the entire composite sample are likely to improve the EMI shielding efficiency. We prepared two batches of the composites using graphene fillers with the distinctively different thicknesses. In the first batch, referred as GF-A, the lateral dimension of the FLG fillers were in the range from $\sim 1.5$ μm to 10 μm and the thicknesses were in the range from 0.35 nm to 12 nm, which corresponds to $1 - 40$ graphene monolayers, respectively. In the second batch, referred to as GF-B, the lateral dimensions were $\sim 2$ μm to 8 μm – almost the same as in the first batch but the thicknesses were much smaller –from 0.35 nm to 3 nm, corresponding to $1 - 8$ graphene monolayers, respectively. The common limitations in the FLG processing technique do not allow one to prepare samples with different thicknesses but exactly the same lateral dimensions[68]. The details of the materials preparation are provided in the Methods section.

An in-house designed mixer was used to disperse the graphene fillers uniformly in the high loading composites[79]. The samples were prepared in the form of disks with the diameter of 25.6 mm and thicknesses from 0.9 mm to 1.0 mm (see Supplementary Table I for the exact thickness of each individual sample). The sample thickness affects the total absorption and the total shielding efficiency of the composites. The optical images of the samples are presented in Figure 1 (a). The samples with the high loading fraction of graphene, $\phi \approx 50 \, \text{wt.\%}$, were characterized by the scanning electron microscopy (SEM). In Figure 1 (b), one can see the overlapping regions of the FLG fillers as well as rolling and bending of the fillers inside the composite. The overlapping of the fillers proves that the graphene loading fraction is above the percolation threshold. It is important to note that while rolling and bending of the fillers may cause an increase in total EMI shielding efficiency of the composites, as a result of increasing internal reflection, it adversely affects the thermal transport properties of the composites[68,72,79]. For this reason, the strategy in materials synthesis was to achieve the electrical and thermal percolation by selecting the right filler dimensions and loading but avoiding the rolling and bending of the fillers.

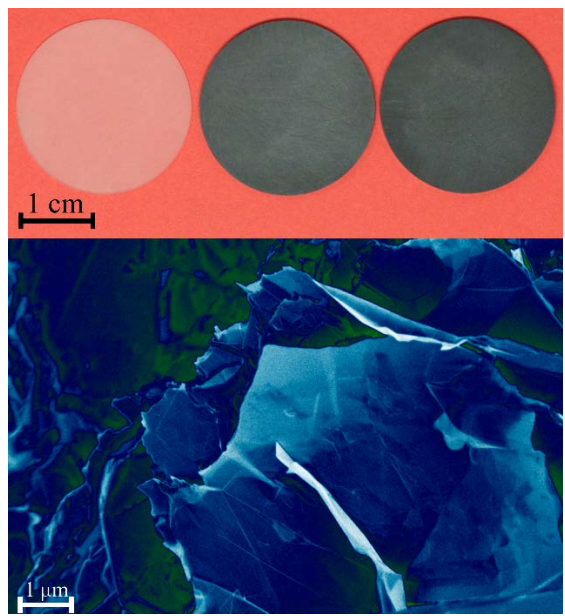

**Figure 1:** (a) From left to right: optical image of pristine epoxy; epoxy with the loading of 50 wt.% of few-layer graphene fillers with the thickness of ~3 nm (8 to 10 layers); epoxy with the loading of 50 wt.% of few-layer graphene fillers with the thickness of ~12 nm (30 to 40 layers). (b) Scanning electron microscopy image of the epoxy with 50 wt% of graphene fillers. The SEM image shows the overlapping of graphene fillers inside the epoxy matrix. Note also the bending and rolling of the fillers at this high-loading fraction composite. The overlapping graphene fillers indicate the formation of a percolation network at high loading fraction of fillers.

## III. Results of Electromagnetic Measurements

The EM scattering parameters, $S_{ij}$, which define the shielding efficiency of the material in terms of reflection ($R$) and transmission ($T$) coefficients were measured using the wave-guide method [1,25,95] (Supplementary Figure S1), with the help of a 2-port programmable network analyzer (PNA, Keysight N5221A) in the frequency range from 8.2 GHz to 12.4 GHz (X-band). Knowing $R$ and $T$, one can calculate the absorption coefficient, $A$, for any incident EM wave as $A = 1 - R - T$. The effective absorption coefficient, $A_{eff} = (1 - R - T)/(1 - R)$ defines the actual absorption characteristic of the EMI shielding material since some part of the incident EM wave energy is reflected at the interface prior to being absorbed or transmitted through it. Experimentally, $S_{ij}$ parameters are measured in decibels (dB) and the subscripts $i$ and $j$ represent the PNA ports which are receiving and sending EM waves, respectively. Therefore, the four scattering parameters $- S_{11}, S_{21}, S_{12}$, and $S_{22} -$ are measured directly by the instrument. In our experiments, port 1 and port 2 are designated to send and receive EM waves to and through the composites, respectively. The reflection and transmission coefficients of the EMI shielding composite can be calculated as $R = |S_{11}|^2 = 10\log(P_R/P_I)$ and $T = |S_{21}|^2 = 10\log(P_T/P_I)$. Here, $P_I$ is the total power of the incident wave on the material, $P_R$ is the reflected power from it, and $P_T$ is the transmitted power through the composite.

Figure 2 (a-c) shows the reflection, absorption and transmission coefficients for pristine epoxy, epoxy with 5 wt%, and epoxy with 50 wt% of GF-A graphene fillers in the X-band frequency. As one can see, for the pristine epoxy, more than 80% of the incident EM wave power is transmitted. Addition of only 5 wt% of GF-A, decreases the transmission coefficient $T$ by almost two times. In this case, most of the incident EM wave power is reflected ($R > 40\%$) at the interface of the EMI shielding material and >10% is absorbed. As the graphene filler concentration increases to $f = 50$ wt%, only 0.002% of the EM wave power is transmitted while the rest is either reflected ($R > 80\%$) or absorbed. It should be noted that most of the incident EM wave is reflected from the surface of the graphene composites. The total shielding efficiency ($SE_{tot}$), which defines the ability of the material to block the incident EM radiation, is the sum of the shielding by reflection, $SE_R = -10\log(1 - R)$, and absorption, $SE_A = -10\log(T/1 - R) = 10\log(1 - A_{eff})$, including multiple-reflections of EM waves within the EMI shielding material[1,2,17,35,96].

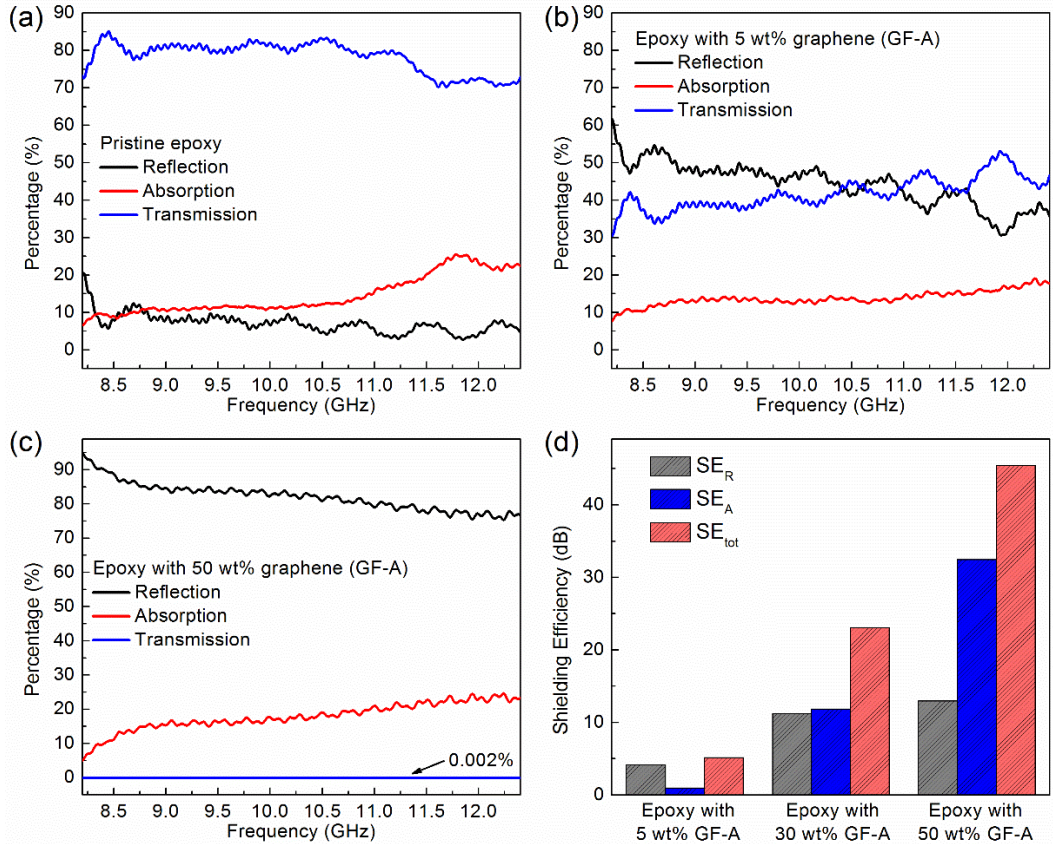

**Figure 2:** Reflection, absorption, and transmission coefficients of (a) pristine epoxy; (b) epoxy with 5 wt% of 12 nm thick graphene (GF-A); and (c) epoxy with 50 wt% of GF-A graphene. At the graphene loading of 50 wt%, only 0.002% of electromagnetic wave power is transmitted through the composite while most of the energy is reflected from the surface. (d) Comparison of the reflection, absorption, and the total shielding efficiency of pristine epoxy, epoxy with 30 wt% GF-A graphene and epoxy with 50 wt% GF-A graphene at the frequency of 8.2 GHz. At the highest loading fraction, the EMI shielding by absorption mechanism overcomes that by the reflection. At the graphene loading of 50 wt%, the total shielding efficiency of the composite exceeds 45 dB.

Figure 2 (d) shows the reflection, absorption and total shielding efficiency of the epoxy with the very low (5 wt%), high (30 wt%) and very high (50 wt%) loading fractions of GF-A graphene fillers. For the epoxy with 5 wt% of graphene, the total shielding efficacy is $SE_{tot} \approx 5$ dB. The $SE_{tot}$ increases with adding more graphene fillers to the polymer matrix, and reaches ~46 dB at 50 wt%. The latter means that more than 99.998% of the incident EM wave is blocked by the composite. Another observation is that increasing the graphene filler loading fraction from 30 wt% to 50 wt%, does not change $SE_R$ significantly. This confirms that most of EM wave power is reflected from the surface of the composite. The shielding by absorption mechanism increases strongly as more electrically conductive fillers are incorporated into the base epoxy matrix. It is interesting to note that composites with graphene fillers below the electrical percolation threshold still reflect and absorb EM wave power. Even though graphene fillers do not form a continuous electrically conductive pass, the incident EM wave can couple to electrons in the individual graphene fillers. This observation confirms that although EMI shielding is strengthened by enhancement of the electric conductivity of the composite, electric percolation is not required for shielding[3]. The fact that graphene composites can block the electromagnetic energy even below the percolation threshold, while remaining electrically insulating, is important for the dual EMI shielding and TIM functionality. Many thermal management applications can only use electrically insulating materials.

Figure 3 (a) shows the reflection shielding efficiency of composites with GF-A graphene fillers as a function of EM wave frequency. One can see that $SE_R$ increases with the filler loading fraction. A weak decrease with increasing frequency of EM wave for each fixed filler loading fraction is also observed. This behavior is in agreement with the so-called Simon formalism[15], where $SE_R$ depends on the electric conductivity of the composite and the frequency of the incident EM wave, according to the expression $SE_R = 50 + 10 \log_{10}(\sigma/f)$. In this equation, $\sigma$ [Scm$^{-1}$] is the electric conductivity and $f$ [MHz] is the frequency of the EM wave. While the reflection shielding efficiency of the pristine epoxy is negligible, it increases abruptly with addition of a small loading fraction of graphene ($\phi < 10$ wt%). For $\phi > 10$ wt%, the increase in $SE_R$ becomes weaker. This trend confirms the saturation of $SE_R$ due to the fact that at $\phi > 10$ wt%, the fillers generated a two-dimensional (2D) network of connected electrically conductive particles on the surface, which exceeds the 2D electrical percolation threshold.

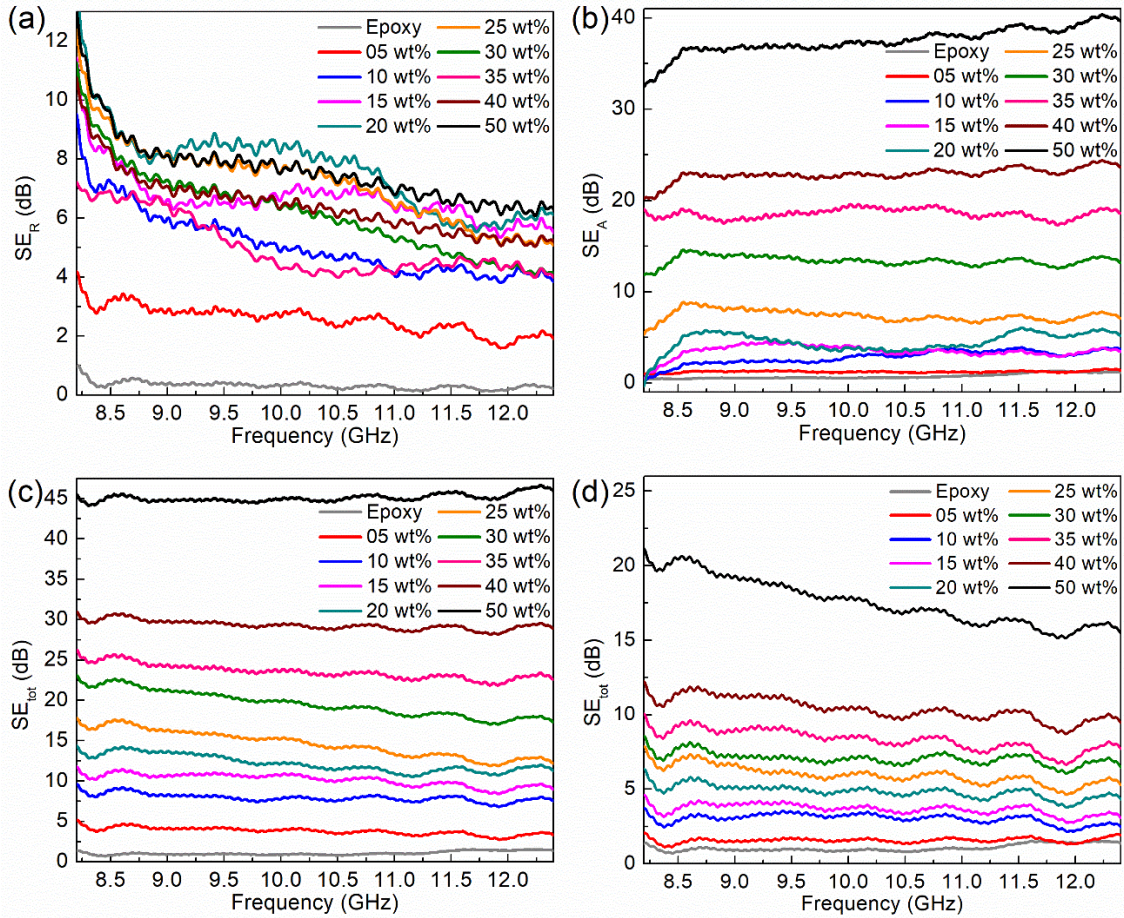

**Figure 3:** (a) Reflection shielding efficiency, $SE_R$ of the epoxy with various loading fractions of GF-A graphene in the frequency range between 8.2 GHz to 12.4 GHz. (b) Absorption shielding efficiency, $SE_A$, of the composites with different loading fraction of GF-A graphene, in the same frequency range. While at lower filler contents the increase in the absorption is gradual, it exhibits an abrupt jump for the loading 20 wt% > $\phi$ > 25 wt%, confirming creation of 3D electrical percolation network. The total shielding efficiency, $SE_{tot}$, of composites with (c) the 12-nm thick GF-A graphene fillers and (d) 3-nm thick GF-B graphene fillers. Although the lateral dimensions of the both fillers are almost the same, the shielding efficiency of composites with the thicker GF-A fillers exceeds that of with GF-B fillers.

In Figure 3 (b) we demonstrate the absorption shielding efficiency of the same composites as a function of the EM wave frequency. With increasing the graphene loading fraction, $SE_A$ increases monotonically. Upon reaching the loading fraction $\phi = 20$ wt. %, the enhancement in the absorption shielding efficiency becomes more pronounced. Similar to the reflection shielding, this trend is attributed to creation of a network of electrically conductive graphene fillers through the in-plane and cross-plane directions of the composite material. The Simon formalism[15] relates $SE_A$ to the thickness, electric conductivity and incident EM wave frequency as $SE_A = 1.7t\sqrt{\sigma f}$ where $t$ [cm] is the thickness of the composite. The formalism does not distinguish between the electrical conductivity through the in-plane and cross-plane directions, assuming a uniform distribution of the conductive filler on the surface – which affects $SE_R$ – and inside the nanocomposite – which affects $SE_A$. The surface in-plane electric resistivity measurements– which affects $SE_R$ in graphene nanocomposites – reported to be ten times lower than the transverse electric resistivity[90]. The results presented in Figures 3 (a-b) demonstrate that the 2D electrical percolation on the surface of the material is achieved before the three-dimensional (3D) volume electrical percolation inside the material. As the composites become thicker, electric percolation may occur at higher filler loading fractions of graphene[90].

Figure 3 (c) shows the total shielding efficiency ($SE_{tot} = SE_R + SE_A$) of the composite with GF-A graphene fillers. According to this plot, $SE_{tot}$ reaches to ~46 dB of the shielding efficiency at the filler concentration of $\phi = 50$ wt%, which exceeds the industry requirements for the EMI shielding materials. We now recall that "graphene" fillers are actually composed of FLG flakes, which can have different average thickness. To confirm the optimum lateral dimensions and thickness of FLG fillers, we repeated the measurements with the epoxy composites filled with the thinner FLG fillers. Figure 3 (d) shows the total shielding efficiency of the epoxy samples with GF-B graphene fillers. The average lateral dimensions, *L*, of GF-A and GF-B graphene filers are similar (*L*~5 μm). However, GF-B consists of graphene fillers with the thickness from ~0.35 nm to ~ 3 nm while GF-A graphene fillers have the thicknesses ranging from ~0.35 nm up to ~12 nm. The thickness of the composite samples in these experiments is 1 mm. The results show that the EM shielding efficiency of the composites with GF-A graphene fillers is almost twice of that of the composites with GF-B fillers, for the same concentration. The reason for better performance of FLG with the intermediate thickness (~0.35 nm up to ~12 nm) is likely related to lesser degradation of their electrical current conducting capabilities upon exposure to matrix material.

Incorporation of SLG to the matrix results in stronger decrease of its electron mobility and electrical conductivity.

## IV. Results of Thermal Conductivity Measurements

As discussed in the introduction, there are two main motivations for creating EMI shielding materials with the high thermal conductivity. From one side, the dual functionality allows such materials to perform both EMI shielding and heat removal. One does not need to use two different materials for these two functions. The latter has important implications for the cost and weight of the system. From the other side, the EMI shielding itself may lead to additional heating of the material. In EMI shielding composites, most of the incident EM wave power is reflected at the interface of the EMI shielding material. However, a significant part of the incident EM power is either absorbed or internally reflected, which results in heat generation inside the EMI shieling material itself. In the high-power EMI shielding applications, if the generated heat is not dissipated efficiently to the environment, it will cause an increase in the temperature and overheating of the EMI material. The increase in temperature adversely affects the electric properties of the composite and, correspondingly, the EMI shielding efficiency. For this reason, the effective EMI shielding material for protection from the high-power EM waves should also have a high thermal conductivity. The latter is often overlooked in the design of EMI shielding materials.

The thermal conductivity of the samples has been measured using the "laser flash technique." We reported our experimental procedures, in the context of other materials, elsewhere[68,71,97,98]. Additional details are provided in the Methods and Supplementary Materials. Figure 4 shows thermal conductivity of the composites with GF-A and GF-B graphene as a function of the graphene filler loading fraction at room temperature. The thermal conductivity of the composite with GF-A graphene fillers increases with increasing loading, reaching the value of $\sim 8\,\mathrm{Wm^{-1}K^{-1}}$ at $\phi = 55$ wt%. This is a substantial, factor of $\times 35$, enhancement in the heat conduction ability as compared to the base matrix material. The composites with the thinner GF-B graphene fillers exhibit lower thermal conductivity enhancement as compared to the composite with GF-A graphene fillers. The latter can be attributed to the fact that thinner fillers can roll and bent easier, which impedes thermal transport through the fillers. The intrinsic thermal conductivity of SLG and thinner FLG can also degrade stronger upon exposure to the matrix material[68]. One should also note here that the synthesized graphene-epoxy composites meet industry standards of the cured TIMs.

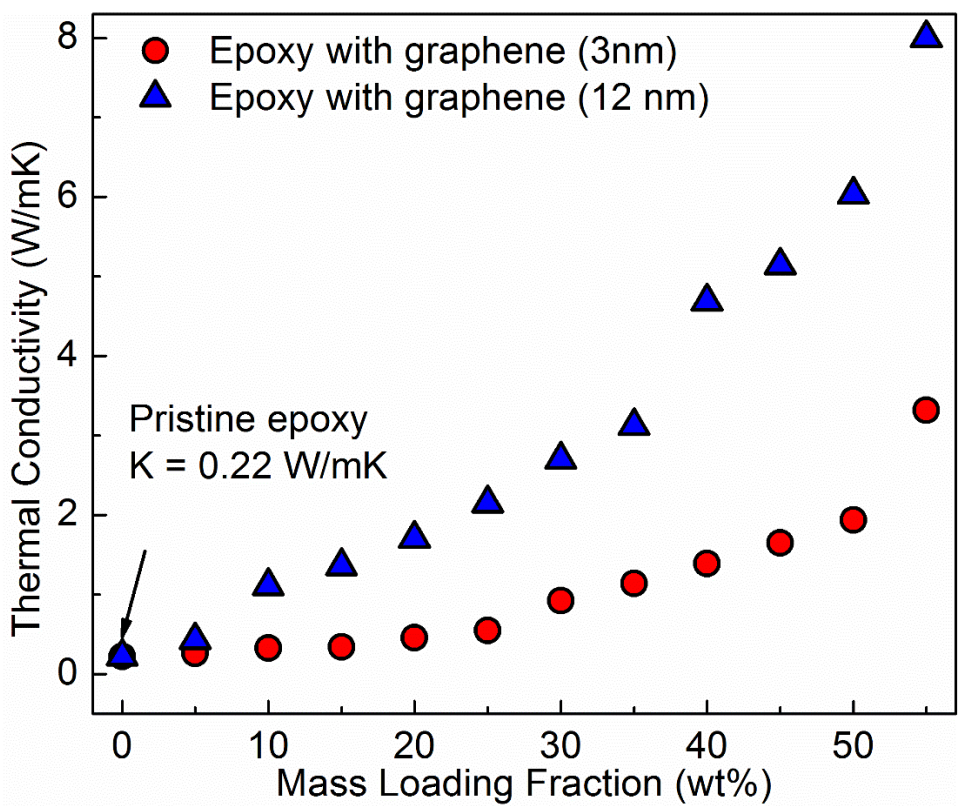

**Figure 4:** Thermal conductivity of the composites with GF-A graphene (blue triangles) and GF-B graphene (red circles) fillers. The thermal conductivity enhancement of the composites with the thicker few-layer graphene fillers is larger than that of the composites with thinner fillers at the same loading fraction. The deviation from linear dependence indicates the on-set of the thermal percolation in the graphene composites.

The deviation from the linear dependence of the thermal conductivity on the loading fraction indicates the thermal percolation threshold at around loading fraction of $\phi = 35$ wt%. The on-set of the thermal percolation happens at somewhat higher loading fractions than the 3D electrical percolation. Thus, optimization of composites for dual EMI shielding and thermal management applications require higher loadings, above the electrical and thermal percolation threshold. However, our results indicate that even low loadings of graphene fillers can improve significantly the EMI shielding and heat conduction properties of polymer composites. We have also established that FLG with the thicknesses in the range from 0.35 nm to 12 nm, which corresponds to $1 - 50$ graphene monolayers, respectively, perform better than composites with thinner FLG. The lateral dimensions were in the few $\mu$m range for both examined cases. Further material synthesis optimization is expected to lead to even better EMI shielding the thermal management properties of graphene composites.

## V. Conclusions

We investigated the EMI shielding efficiency and thermal conductivity of composites with graphene. It was found that composites with the few-layer graphene fillers reveal an efficient total electromagnetic interference shielding, $SE_{tot} \approx 45$ dB, in the X-band frequency range while simultaneously providing the high thermal conductivity, $K \approx 8\,\mathrm{Wm^{-1}K^{-1}}$, which is a factor of $\times 35$ larger than that of the base matrix material. Our results show that graphene composites can block more than 99.998% of the high-frequency EM radiation while providing an important function of heat removal. These properties allow for a dual functional application of graphene composites: EMI shielding and thermal management. The loading fractions of graphene above the electrical and thermal percolation threshold allow one to strongly enhance both the EMI shielding and heat conduction properties. From the other side, we established that graphene composites can efficiently block the electromagnetic energy even below the percolation threshold while remaining *electrically insulating*. The latter is an important feature for some TIM applications. The dual functionality of the graphene composites can substantially improve the EMI shielding and thermal management of the airborne systems while simultaneously reducing their weight and cost.

## METHODS

**Sample Preparation:** The composite samples were prepared by mixing the commercially available FLF flakes (Graphene Supermarket) with epoxy (Allied High Tech Products, Inc.). For samples with loading fraction, the epoxy resin and the filler were mixed using a high-shear speed mixer (Flacktek Inc.) at 800 rpm and 2000 rpm each for 5 minutes. The mixture was vacuumed for 30 minutes. After that time, the curing agent (Allied High Tech Products, Inc.) was added in the mass ratio of 12:100 with respect to the epoxy resin. The mixture was mixed and vacuumed one more time and left in the oven for ~2 hours at 70º C in order to cure and solidify. For very high loading samples, graphene was added to the resin at three different steps and dispersed using the mixture at 2000 rpm for 5 minutes each time. The mixture was vacuumed for 5 minutes. Then, the curing agent was added, and the solution was mixed at high rotation speeds of 3500 rpm and 2000 rpm for 15 seconds and 10 minutes, respectively. The homogenous mixture was gently pressed and left in the oven at 70 ºC for ~2 hours to cure. The details of the sample preparation can be found in the Supplementary Materials.

**Electromagnetic Interference Shielding Measurements:** EMI measurements were performed in 8.2 – 12.4 GHz frequency range with frequency resolution of 3 MHz and input power of 3 dBm using a Programmable Network Analyzer (PNA) Keysight N5221A. As a sample holder, we used WR-90 commercial grade straight waveguide with two adapters at both ends with SMA coaxial ports. The samples with diameter $d \geq 25$ mm were a bit larger than the rectangular cross section (22.8×10.1 mm$^2$) of the central hollow part of the waveguide in order to prevent the leakage of EM waves from the sender to receiver antenna. The scattering parameters, $S_{ij}$, were directly measured and used to extract the reflection and absorption shielding efficiency of the composites. More details on the experimental setup and procedures can be found in the Supplementary Materials.

**Thermal Conductivity Measurements:** Thermal conductivity of the samples were measured using the transient "laser flash" technique (LFA 467 HyperFlash, Netzsch) compliant with the international standards of ASTM E-1461, DIM EN 821 and DIN 30905. Using this technique, we measured the thermal diffusivity ($\alpha$) of the samples, which, in turn, was used to determine the thermal conductivity of the composites according to the equation $K = \rho \alpha c_p$ where $K$, $\rho$, and $c_p$ are the thermal conductivity, density, and specific heat, respectively. In LFA technique, a Xenon flash lamp introduces an energy pulse to one side of the sample. The time dependent temperature rise on the opposite side of the sample was measured by an infrared detector. The thermal diffusivity was then extracted by calculating the time constant of temperature rise. More details on the LFA thermal conductivity method can be found in Supplementary Materials.

### *Acknowledgements*

This work was supported, in part, by NSF through the Emerging Frontiers of Research Initiative (EFRI) 2-DARE award EFRI-1433395: Novel Switching Phenomena in Atomic $MX_2$ Heterostructures for Multifunctional Applications, and by the UC-National Laboratory Collaborative Research and Training Program - University of California Research Initiatives LFR-17-477237.

### Contributions

A.A.B. and F.K. conceived the idea of the study. A.A.B. coordinated the project and contributed to the experimental and theoretical data analysis; F.K. performed EM shielding measurements and conducted data analysis; Z.B. prepared the composites, performed thermal and EM shielding measurements and assisted with the data analysis; M.B. contributed to EM shielding measurements; A.S.M. assisted with sample preparation and contributed to thermal conductivity measurements; J.S.L. conducted materials characterization. A.A.B. led the manuscript preparation. All authors contributed to writing and editing of the manuscript. The authors thank Dr. A. Khitun (UC Riverside) for providing electromagnetic equipment and Alec Balandin (Riverside STEM Academy) for assistance with the composite preparations.